\documentclass[12pt, fleqn]{article}
\usepackage[cp1251]{inputenc}
\usepackage{latexsym,amsfonts,amssymb}
\usepackage{graphicx}

\usepackage{amsbsy}
\usepackage{amsmath}
\usepackage{epsf}
\usepackage{cite}

\newtheorem{theo}{Theorem}
\newtheorem{remark}{Remark}

\newcommand{\bt}{\begin{theo}}
\newcommand{\et}{\end{theo}}
\newcommand{\bd}{\begin{displaymath}}
\newcommand{\ed}{\end{displaymath}}

\newcommand{\lf}{\left}
\newcommand{\rg}{\right}

\newcommand{\be} {\begin{equation}}
\newcommand{\ee} {\end{equation}}
\newcommand{\ba}{\begin{array}}
\newcommand{\ea} {\end{array}}
\newcommand{\bea}{\begin{eqnarray}}
\newcommand{\eea} {\end{eqnarray}}

\newcommand{\p} {\partial}

\begin{document}

\begin{center}
 {\Large \bf Conditional symmetries and  exact solutions \\ 
 of a nonlinear three-component reaction-diffusion model. }
\medskip

{\bf Roman Cherniha,$^{a}$\footnote{\small   E-mail: r.m.cherniha@gmail.com}}
  {\bf Vasyl' Davydovych$^a$}\footnote{\small E-mail: davydovych@imath.kiev.ua}
 \\
{\it $^a$~Institute of Mathematics,  National Academy
of Sciences  of Ukraine,\\
 3, Tereshchenkivs'ka Street, 01601 Kyiv, Ukraine
}\\
 \end{center}

\begin{abstract}
Q-conditional (nonclassical) symmetries of the known three-component reaction-diffusion system
[K. Aoki et al Theor. Pop. Biol. 50(1) (1996)] modeling  interaction between  farmers and hunter-gatherers
are constructed for the first time. A wide variety of Q-conditional symmetries are found in an explicit form and it is shown that these symmetries are  not equivalent to the Lie symmetries.  Some operators of Q-conditional (nonclassical) symmetry are applied for finding exact solutions of the reaction-diffusion system in question. Properties   of the exact solutions (in particular, their  asymptotic behaviour)  are identified and possible biological interpretation is discussed.

\end{abstract}

\textbf{Keywords:} reaction-diffusion system;
hunter-gatherer--farmer system; Lie symmetry;\\ $Q$-conditional symmetry; exact solution.

\section{\bf  Introduction} \label{sec:1}

In
\cite{ao-sh-shige-96}, a three-component model
 for describing the spread of an initially
localized population of farmers into a
 region occupied by hunter-gatherers  was introduced. Under some assumptions clearly indicated
 in \cite{ao-sh-shige-96}, the spread and interaction between  farmers and hunter-gatherers
  can be modeled as a reaction-diffusion (RD) process in the form of the three-component system of nonlinear PDEs . Recently, the model was used for mathematical  description of some other phenomena. For example, a model   describing language competition was derived in \cite{straughan-2014}. The model is based on the  three-component system of nonlinear PDEs, which has the same structure as the system introduced in \cite{ao-sh-shige-96}, however some coefficients have opposite signs.
  Notably, the model proposed in \cite{straughan-2014} is a modification of another model for language competition developed earlier  in  \cite{kandler-2010} (see also \cite{kandler-18}).

  Here we study the original model from the paper \cite{ao-sh-shige-96} used for modeling competition between farmers and hunter-gatherers. The work is a natural continuation of  our recent paper  \cite{ch-dav-2017}, in which Lie symmetries and  traveling  fronts  of this model have been studied.
  After the relevant re-scaling (see \cite{ch-dav-2017} for details), the model takes the form of the  nonlinear RD system
\begin{equation}\label{1-2}\begin{array}{l}  u_t = d_1 u_{xx}+u(1-u-a_1v),\\  v_t =
d_2 v_{xx}+ a_2v(1-u-a_1v)+uw+a_1vw,\\  w_t = d_3
w_{xx}+a_3w(1-w)-a_4uw-a_5vw,  \end{array}\end{equation}
where $u(t,x), \ v(t,x)$ and $w(t,x)$ are non-dimensional densities of the three
populations of initial farmers, converted farmers, and
hunter-gatherers, respectively (hereafter the lower subscripts $t$ and
$x$  mean differentiation w.r.t. these variables).
Here
(\ref{1-2}) is called the hunter-gatherer--farmer (HGF) system and
one is the main object of investigation in this paper. We naturally
assume that the diffusivities $d_1,
\ d_2$ and $d_3$ are positive constants. Other parameters are non-negative constant, excepting $a_4$  that is a positive  constant (otherwise the carrying
capacity of farmers is zero\cite{ch-dav-2017}).  Obviously, the HGF system (\ref{1-2})
is not a particular case of   the well-known diffusive Lotka--Volterra (DLV) system
\be\label{1-1*}\ba{l}  u_t =  d_1u_{xx}+u(a_1+b_1u+c_1v+d_1w),\\
 v_t = d_2v_{xx}+ v(a_2+b_2u+c_2v+d_2w),\\  w_t =
d_3w_{xx}+w(a_3+b_3u+c_3v+d_3w),
\ea\ee
because of the term $uw$ in the second equation of (\ref{1-2}). Notably,
in the special case $d_1=d_2$,  $a_1\neq0$ and $a_2=1$, system (\ref{1-2}) is
reduced to the DLV system by the transformation $u+a_1v \rightarrow
v.$

In contrast to our previous study \cite{ch-dav-2017}, which is devoted to  Lie symmetries, here we search for  $Q$-conditional symmetry
 (nonclassical symmetry) of the HGF system (\ref{1-2}).
 It is  well-known that  the notion of nonclassical symmetry was introduced in  \cite{1-bl-c} and plays
an important role in investigation of nonlinear PDEs
(see, review  \cite{1-saccomandi05} and monographs
\cite{bl-ch-anco, ch-dav-book, ch-se-pl-book} for more details).  In particular,
having such symmetries in an explicit form, one may construct new
exact solutions, which are not obtainable  by the classical Lie
algorithm.

 The algorithm for finding $Q$-conditional symmetry (following \cite{1-f-s-s}, we  use this terminology  instead of
nonclassical symmetry) of a given PDEs is based on the classical Lie method \cite{ovs80, olv86}. However, in contrast to the case of Lie symmetry, the corresponding
  system of determining equations (DEs) is {\it nonlinear} and its general solution can be found only in exceptional cases.  So, obtaining an
exhaustive description
 of   $Q$-conditional symmetry of the given equation
 is a non-trivial and   difficult task.  As a result, scalar PDEs only were under study for a long time (see extensive reviews about this matter in Chapter 1 of \cite{ch-dav-book}) because systems of DEs for systems of PDEs are much more complicated.
To the best of our  knowledge,  there are  only a few  papers
devoted to the  search of $Q$-conditional symmetries for  systems of PDEs
published before 2010 \cite{all-nucci-96,
bar2, ch-se-2003, ch-pl-08, murata-06}.
 A majority of such papers were published during the
current  decade \cite{2-arrigo2010, ch-2010, ch-dav-2011,ch-dav-2013,torissi,ch-dav-2015}.  It should be stressed that only the
papers \cite{all-nucci-96} (see section 4.1) and  \cite{ch-dav-2013}
are devoted  to construct   $Q$-conditional symmetry
  operators for a \emph{three-component PDE system} while two-component systems only are under study in other papers. Notably, the conditional symmetries of the three-component Prandtl system derived in \cite{all-nucci-96} coincide with the relevant  Lie symmetries. In paper \cite{ch-dav-2013}, some
  $Q$-conditional symmetries of the three-component DLV system are found  and it is shown that they are not obtainable by the classical Lie method.
  Here we make essential progress comparing with the papers cited above because {\it all possible $Q$-conditional symmetries}
  of the HGF system (\ref{1-2}) are constructed in an explicit form.

  It should be also mentioned that there is a further generalization of the notion of $Q$-conditional symmetry -- generalized conditional symmetry -- introduced in \cite{Foka94} (the terminology `conditional Lie-B\"acklund symmetry' suggested in \cite{zh95} has the same definition but one is rather misleading). Recently some two-component systems of evolution equations (including reaction-diffusion equations) have been studied using the generalized conditional symmetry method \cite{ji-qu-2014},\cite{wang-ji-2015}.

The paper is organized as follows. In Section~\ref{sec:2}, the main theorem  about
$Q$-conditional symmetries of the HGF system (\ref{1-2}) is proved. In Section~\ref{sec:3}, the most interesting (from
applicability point of view) case of system (\ref{1-2}) is examined.
In particular,  non-Lie ans\"{a}tze are derived and applied for
reducing the system in question to systems of ODEs. The reduced
systems are analyzed in order to construct exact solutions, some highly nontrivial exact solutions are derived and their properties
are studied. Finally,
we briefly discuss the result obtained and present some conclusions
in the last section.

\section{\bf $Q$-conditional symmetries of the HGF system (\ref{1-2})}\label{sec:2}

First of all, we remind the reader that
similarly to  Lie symmetries,  $Q$-conditional symmetries
 are constructed in the form of  the
first-order differential operators \be\label{1-6*}\ba{l} \hskip1cm Q
= \xi^0
(t, x, u, v,w)\p_{t} + \xi^1 (t, x, u, v,w)\p_{x} +\\
 \eta^1(t, x, u, v, w)\p_{u}+\eta^2(t, x, u, v, w)\p_{v}+\eta^3(t, x, u, v,
 w)\p_{w}, \ (\xi^0)^2+(\xi^1)^2\neq0,\ea\ee where the  coefficients
$\xi^0,\ \xi^1$ and $\eta^k$ ($k=1,2,3$) should be found using the
well-known criterion. Taking into account the property of the
$Q$-conditional symmetries (operator can be multiplied by an
arbitrary smooth function), operator (\ref{1-6*}) has essentially
 different forms in the cases $\xi^0\neq0$ and $\xi^0=0$, namely
\be\label{1-6} Q = \p_{t} + \xi(t, x, u, v,w)\p_{x} +
 \eta^1(t, x, u, v, w)\p_{u}+\eta^2(t, x, u, v, w)\p_{v}+\eta^3(t, x, u, v, w)\p_{w},  \ee
and \be\label{1-6**} Q = \p_{x} +
 \eta^1(t, x, u, v, w)\p_{u}+\eta^2(t, x, u, v, w)\p_{v}+\eta^3(t, x, u, v, w)\p_{w}.\ee
It turns out that the  systems of DEs for finding $Q$-conditional
symmetries (\ref{1-6}) and  (\ref{1-6**}) are essentially different
(see, e.g., \cite[Section 1.4]{ch-dav-book}
). Hereafter we concentrate ourselves   on the case~$\xi^0\neq0.$

According to the standard criteria (see, e.g., \cite[Section
2.2]{ch-dav-book}), operator (\ref{1-6}) is the $Q$-conditional
symmetry (non-classical symmetry) for the HGF system (\ref{1-2}) if
the following invariance conditions are satisfied:
\be\label{1-7}
\begin{array}{l}
\mbox{\raisebox{-1.6ex}{$\stackrel{\displaystyle  
Q}{\scriptstyle 2}$}} \lf(S_1\rg)\Big\vert_{{\cal{M}}}  
 \equiv\mbox{\raisebox{-1.6ex}{$\stackrel{\displaystyle  
Q}{\scriptstyle 2}$}}\,  
\big(d_1 u_{xx}-u_t+u(1-u-a_1v)\big)  
\Big\vert_{{\cal{M}}}=0, \\[0.3cm]  
\mbox{\raisebox{-1.6ex}{$\stackrel{\displaystyle  
Q}{\scriptstyle 2}$}} \lf(S_2\rg)\Big\vert_{{\cal{M}}}  
 \equiv\mbox{\raisebox{-1.6ex}{$\stackrel{\displaystyle  
Q}{\scriptstyle 2}$}}\,  
\big(d_2 v_{xx}-v_t+ a_2v(1-u-a_1v)+uw+a_1vw\big)  
\Big\vert_{{\cal{M}}}=0, \\[0.3cm]  
\mbox{\raisebox{-1.6ex}{$\stackrel{\displaystyle  
Q}{\scriptstyle 2}$}} \lf(S_3\rg)\Big\vert_{{\cal{M}}}  
 \equiv\mbox{\raisebox{-1.6ex}{$\stackrel{\displaystyle  
Q}{\scriptstyle 2}$}}\,  
\big(d_3
w_{xx}-w_t+a_3w(1-w)-a_4uw-a_5vw\big)  
\Big\vert_{{\cal{M}}}=0,
\end{array}  \ee  where  operator $ \mbox{\raisebox{-1.6ex}{$\stackrel{\displaystyle  
Q}{\scriptstyle 2}$}} $  
is the second  
 prolongation of the operator $Q$, the manifold \[{\cal{M}} = \{S_1=0,\ S_2=0,\ S_3=0,\
 Q(u)=0,\
Q(v)=0,\ Q(w)=0\}.\]

The second prolongation of the operator $Q$ has the form
\[ \begin{array}{l} \medskip
\mbox{\raisebox{-1.6ex}{$\stackrel{\displaystyle  
Q}{\scriptstyle 2}$}}
=Q+\rho^1_t\frac{\p}{\p u_t}+\rho^2_t\frac{\p}{\p v_t}+\rho^3_t\frac{\p}{\p w_t}+
\rho^1_x\frac{\p}{\p u_x}+\rho^2_x\frac{\p}{\p v_x}+\rho^3_x\frac{\p}{\p w_x}+
\sigma^1_{tx}\frac{\p}{\p u_{tx}}+ \sigma^2_{tx}\frac{\p}{\p
v_{tx}} + \sigma^3_{tx}\frac{\p}{\p w_{tx}}+\\ \medskip
\qquad\sigma^1_{tt}\frac{\p}{\p
u_{tt}}+\sigma^2_{tt}\frac{\p}{\p v_{tt}}+\sigma^3_{tt}\frac{\p}{\p
w_{tt}} +\sigma^1_{xx}\frac{\p}{\p
u_{xx}}+\sigma^2_{xx}\frac{\p}{\p v_{xx}}+\sigma^3_{xx}\frac{\p}{\p
w_{xx}}, \end{array} \]  where  the coefficients $\rho^k$ and
$\sigma^k$ ($k=1,2,3$) with the relevant indices are calculated by
the well-known formulae (see, e.g., \cite{ch-dav-book,olv86,ovs80}).

Now we apply the rather standard procedure for obtaining system of
DEs, using the invariance conditions (\ref{1-7}). From the formal
point of view, the procedure is the same as for Lie symmetry search,
however, six (not three\,!) different derivatives, say $u_{xx}, \
v_{xx}, \ w_{xx},\ u_t, \ v_t$ and $w_t$,  can be excluded using the
manifold ${\cal{M}}$. After straightforward calculations, one
arrives~at~the \emph{nonlinear system} of DEs
\bea && \nonumber 1)\ \medskip
\xi_{u}=\xi_{v}=\xi_{w}=0,
\\ && \nonumber 2)\ \medskip
\eta^k_{uu}=\eta^k_{uv}=\eta^k_{uw}=\eta^k_{vv}=\eta^k_{vw}=\eta^k_{ww}=0,
\ k=1,2,3,
\\ && \nonumber 3)\ \medskip (d_1-d_2)\,\xi\eta^1_v-2d_1d_2\eta^1_{xv}=0,
\quad  (d_1-d_3)\,\xi\eta^1_w-2d_1d_3\eta^1_{xw}=0, \\
 && \nonumber 4)\ \medskip (d_1-d_2)\,\xi\eta^2_u+2d_1d_2\eta^2_{xu}=0,
\quad (d_2-d_3)\,\xi\eta^2_w-2d_2d_3\eta^2_{xw}=0,
\\ && \nonumber 5)\ \medskip (d_1-d_3)\,\xi\eta^3_u+2d_1d_3\eta^3_{xu}=0,
\quad (d_2-d_3)\,\xi\eta^3_v+2d_2d_3\eta^3_{xv}=0,
\\ && \nonumber 6)\  \medskip\xi_t-d_1\xi_{xx}+2d_1\eta^1_{xu}+2\xi\xi_x=0,
\\ && \nonumber 7)\ \medskip \xi_t-d_2\xi_{xx}+2d_2\eta^2_{xv}+2\xi\xi_x=0,
\\ && 8)\ \medskip \xi_t-d_3\xi_{xx}+2d_3\eta^3_{xw}+2\xi\xi_x=0, \label{1-3}
\eea \bea && \nonumber 9)\
\eta^1C^1_u+\eta^2C^1_v+\eta^3C^1_w+\lf(2\xi_x-\eta^1_u\rg)C^1-\frac{d_1}{d_2}\,\eta^1_vC^2-
\frac{d_1}{d_3}\,\eta^1_wC^3+\\ && \nonumber \medskip \hskip0.5cm
d_1\eta^1_{xx}-\eta^1_t-2\xi_x\eta^1+\lf(\frac{d_1}{d_2}-1\rg)\eta^2\eta^1_v+
\lf(\frac{d_1}{d_3}-1\rg)\eta^3\eta^1_w=0,\\ && \nonumber
10)\
\eta^1C^2_u+\eta^2C^2_v+\eta^3C^2_w+\lf(2\xi_x-\eta^2_v\rg)C^2-\frac{d_2}{d_1}\,\eta^2_uC^1-
\frac{d_2}{d_3}\,\eta^2_wC^3+\\ && \nonumber \medskip \hskip0.5cm
d_2\eta^2_{xx}-\eta^2_t-2\xi_x\eta^2+\lf(\frac{d_2}{d_1}-1\rg)\eta^1\eta^2_u+
\lf(\frac{d_2}{d_3}-1\rg)\eta^3\eta^2_w=0,\\ && \nonumber
11)\
\eta^1C^3_u+\eta^2C^3_v+\eta^3C^3_w+\lf(2\xi_x-\eta^3_w\rg)C^3-\frac{d_3}{d_1}\,\eta^3_uC^1-
\frac{d_3}{d_2}\,\eta^3_vC^2+\\ && \nonumber  \hskip0.5cm
d_3\eta^1_{xx}-\eta^3_t-2\xi_x\eta^3+\lf(\frac{d_3}{d_1}-1\rg)\eta^1\eta^3_u+
\lf(\frac{d_3}{d_2}-1\rg)\eta^2\eta^3_v=0,
 \eea
 where \be\label{1-8}\begin{array}{l}
 C^1=u(1-u-a_1v),\\ C^2=a_2v(1-u-a_1v)+uw+a_1vw,\\
 C^3=a_3w(1-w)-a_4uw-a_5vw.
 \end{array}\ee

\begin{remark} The nonlinear system (\ref{1-3}) is the system of DEs
for the general RD system of the form
\begin{equation}\nonumber\begin{array}{l}  u_t = d_1 u_{xx}+C^1(u,v,w),\\  v_t =
d_2 v_{xx}+ C^2(u,v,w),\\  w_t = d_3 w_{xx}+C^3(u,v,w),
\end{array}\end{equation} where $C^k \ (k=1,2,3)$ are arbitrary
smooth functions and $d_k>0.$ Here we examine this system in the case when the functions $C^k \ (k=1,2,3)$ possess the form  (\ref{1-8}).
 \end{remark}

Solving the linear  subsystem   $1)$--$2)$ of system (\ref{1-3}),
one specifies the form of operator (\ref{1-6}) as follows
\be\label{1-5}\begin{array}{l} \medskip  Q=\p_t+\xi(t,x)\p_x+
\big(r^1(t,x)u+q^1(t,x)v+h^1(t,x)w+p^1(t,x)\big)\p_u+\\ \medskip
\hskip1cm
\big(r^2(t,x)v+q^2(t,x)u+h^2(t,x)w+p^2(t,x)\big)\p_v+\\ \hskip1cm
\big(r^3(t,x)w+q^3(t,x)u+h^3(t,x)v+p^3(t,x)\big)\p_w, \end{array}\ee
where   $\xi, \ r^k, \ q^k, \ h^k$  and  $p^k$  are unknown smooth functions
at the moment. So, any  operator of conditional symmetry are linear w.r.t.  $u, \ v$  and $w$.

Now we formulate the main result of this section.

\bt\label{th-1} System (\ref{1-2})
     is invariant under $Q$-conditional symmetry operator(s) of the form (\ref{1-5}) if and only if one  and the
corresponding operator(s) have the forms listed in Table~\ref{tab2}.
  \et

\begin{table}
\caption{$Q$-conditional symmetry operators of system  (\ref{1-2})}
\medskip
\label{tab2}       
\begin{tabular}{p{0.2cm}p{3.6cm}p{2.5cm}p{8.4cm}}
\hline\noalign{\smallskip} & Reaction terms &Restrictions & Operator(s)  \\ \hline &&&\\
 1 &
$u(1-u)$  \newline{$\frac{d_2}{d_1}\,v(1-u)+uw$}
\newline{$a_3w(1-w)-a_4uw$}
 & $a_3\neq0$& $\p_t+\mu\p_x+\big(q^2u+p^2\big)\p_v,$
 \newline{$q^2=\lf(\alpha_1+\alpha_2e^{\frac{d_2}{d_1}\,t}\rg)E(d_1,d_2),$}
\newline{$p^2=-\alpha_2e^{\frac{d_2}{d_1}\,t}E(d_1,d_2)$}  \\ \hline &&&\\
 2 & $u(1-u-a_1v)$ \newline{$\frac{d_2-d_3}{d_1-d_3}\,v(1-u-a_1v)+$}\newline{$uw+a_1vw$}
\newline{$-\frac{d_3}{d_1}uw-\frac{a_1d_3}{d_1}vw$}  & $a_1\neq0$ \newline{$d_1\neq d_2$}
\newline{$d_1\neq d_3$} & $\p_t-\frac{d_1}{d_1-d_2}\,w(a_1\p_u-\p_v)-
\frac{d_3}{d_1-d_3}\,w\p_w;$
\newline{$\p_t-\frac{d_3}{a_1(d_1-d_3)}\,u\left(a_1\p_u-\p_v\right)-$}\newline{$
\frac{(d_1-d_2)d_3^2}{a_1d_1(d_1-d_3)^2}\,u\p_w$}\\ \hline &&&\\
3 & $u(1-u-a_1v)$
\newline{$\frac{d_2}{d_1}\,v(1-u-a_1v)+$}\newline{$uw+a_1vw$}
\newline{$-a_4uw-a_1a_4vw$}  & $a_1\neq0$ \newline{$d_1\neq d_2$}
\newline{$a_4d_1\neq d_3$} & $(d_1-d_2)\p_t+$\newline{$
\lf(\frac{d_2}{a_1}(u-1)+d_2v+\frac{d_1d_3}{a_4d_1-d_3}\,w\rg)\left(a_1\p_u-\p_v\right)$}
\\ \hline &&&\\
4 & $u(1-u-a_1v)$
\newline{$v(1-u-a_1v)+$}\newline{$uw+a_1vw$}
\newline{$-a_4uw-a_1a_4vw$}  & $a_1\neq0$ \newline{$d_1=d_2=d$}
\newline{$a_4d\neq d_3$} &
$\p_t+\Big(\big(\alpha_1+\alpha_2(a_4d-d_3)\,e^t\big)u+$\newline{$
\alpha_2a_1(a_4d-d_3)\,e^tv+\alpha_2a_1
d_3\,e^tw+$}\newline{$\alpha_2(d_3-a_4d)\,e^t\Big)\left(a_1\p_u-\p_v\right)$}
\\ \hline &&&\\
5 & $u(1-u-a_1v)$
\newline{$v(1-u-a_1v)+$}\newline{$uw+a_1vw$}
\newline{$-\frac{d_3}{d}\,uw-a_1\frac{d_3}{d}\,vw$}  & $a_1\neq0$ \newline{$d_1=d_2=d$}
 &
$\p_t+\mu\p_x+$\newline{$\big(\alpha_1u+ \alpha_2e^tE(d_3,d)
w\big)\left(a_1\p_u-\p_v\right)$}
\\ \hline&&&\\
 6 &
$u(1-u)$  \newline{$a_2v(1-u)+uw$}
\newline{$-a_4uw$}
 & $a_4d_2\neq a_2d_3$ \newline{$a_2d_1\neq d_2$} \newline{$d_2\neq d_3$}
  & $(d_2-d_3)\p_t+\Big(\alpha_1v+\frac{d_3(a_2d_3+\alpha_1)}{a_4d_2-a_2d_3}\,w+
  $ \newline{$P(t,x)\Big)\p_v-a_2d_3\,w\p_w,  \ a_2P(t,x)=0$}
\\
\hline &&&\\
 7 &
$u(1-u)$  \newline{$\frac{a_4d_2}{d_3}\,v(1-u)+uw$}
\newline{$-a_4uw$}
 & & $\p_t+\mu\p_x+\big(\alpha_1v+h^2w\big)\p_v+\alpha_1w\p_w,$
 \newline{$h^2=\alpha_2\exp\lf(\frac{a_4d_2+\alpha_1(d_2-d_3)}{d_3}\,t\rg)E(d_3,d_2)$}  \\
\hline
\end{tabular}
\end{table}

\begin{table} 
\begin{tabular}{p{0.2cm}p{3.6cm}p{2.5cm}p{8.4cm}}
\hline\noalign{\smallskip} & Reaction terms &Restrictions & Operator(s)  \\ \hline  &&&\\
  8 &
$u(1-u)$  \newline{$\frac{d_2}{d_1}\,v(1-u)+uw$}
\newline{$-a_4uw$}
 &$a_4d_1\neq d_3$& $\p_t+\mu\p_x+\big(\alpha_1v+q^2u+p^2\big)\p_v+\alpha_1w\p_w,$
 \newline{$q^2=\lf(\alpha_2+\alpha_3e^{\frac{d_2}{d_1}\,t}\rg)E(d_1,d_2),$}
\newline{$p^2=-\alpha_3e^{\frac{d_2}{d_1}\,t}E(d_1,d_2);$}
\newline{$(d_2-d_3)\p_t+\lf(\alpha_1
v+\frac{d_3(d_2d_3+d_1\alpha_1)}{d_2(a_4d_1-d_3)}\,w+\rg.$}
\newline{$\lf.\alpha_2u+\alpha_3e^{\frac{d_2}{d_1}\,t}(u-1)\rg)\p_v-\frac{d_2d_3}{d_1}\,w\p_w$}\\
\hline&&&\\
  9 &
$u(1-u)$  \newline{$\frac{d_2}{d_1}\,v(1-u)+uw$}
\newline{$-\frac{d_3}{d_1}uw$}
 && $\p_t+\mu\p_x+\big(\alpha_1v+q^2u+h^2w+p^2\big)\p_v+\alpha_1w\p_w,$ \newline{$p^2=-\alpha_3e^{\frac{d_2}{d_1}\,t}E(d_1,d_2),$}
 \newline{$q^2=\lf(\alpha_2+\alpha_3e^{\frac{d_2}{d_1}\,t}\rg)E(d_1,d_2),$}
\newline{$h^2=\alpha_4\exp\lf(\frac{d_2d_3+\alpha_1d_1(d_2-d_3)}{d_1d_3}\,t\rg)E(d_3,d_2)$}  \\
\hline&&&\\
  10 &
$u(1-u)$  \newline{$\frac{d_2-d_3}{d_1}\,v(1-u)+uw$}
\newline{$-\frac{d_3}{d_1}uw$}
 &$d_2\neq d_3$& $\p_t+\big(\alpha_1v+q^2u\big)\p_v+\big(\alpha_1w+q^3u+\alpha_2\big)\p_w,$
 \newline{$q^3=\alpha_3e^{-\frac{d_3}{d_1}\,t}-\alpha_2, \ q^2=-\frac{d_1}{d_3}q^3;$}
\newline{$\p_t+\lf(\alpha_1
v+\frac{\alpha_1d_1+d_3}{d_3}\,w\rg)\p_v-\frac{d_3}{d_1}\,w\p_w$}
 \\
\hline&&&\\
  11 &
$u(1-u)$  \newline{$a_2v(1-u)+uw$}
\newline{$-\frac{d_3}{d_1}uw$}
 &$ a_2d_1\neq d_2$ \newline{$a_2d_1\neq d_2-d_3$}\newline{$d_2\neq d_3$} \newline{$a_2P(t,x)=0$}&
 $\p_t+\lf(\alpha_1v-\frac{d_1\alpha_2}{a_2d_1-d_2}\,u+
 P(t,x)\rg)\p_v+$\newline{$\big(\alpha_2\lf(1-u\rg)+\alpha_1w\big)\p_w;
 (d_2-d_3)\p_t-a_2d_3w\p_w+$}\newline{$\lf(\alpha_1
v+\frac{d_1(a_2d_3+\alpha_1)}{d_2-a_2d_1}\,w+P(t,x)\rg)\p_v;$}
\newline{$\p_t+\lf(-\frac{d_3}{d_1}\,v+\frac{\alpha_1
d_1}{a_2d_1-d_2}e^{-\frac{d_3}{d_1}\,t}u+\rg.$}\newline{$\lf.
\frac{d_3(a_2d_1-d_2+d_3)}{(a_2d_1-d_2)(d_3-d_2)}\,w+P(t,x)\rg)\p_v+$}\newline{$
\lf(\frac{a_2d_3}{d_3-d_2}\,w+\alpha_1
e^{-\frac{d_3}{d_1}\,t}u\rg)\p_w$}
 \\
\hline &&&\\
   12 &
$u(1-u)$  \newline{$a_2v(1-u)+uw$}
\newline{$-\frac{d}{d_1}uw$}
 &$ a_2\neq 0$ \newline{$d_2=d_3=d$} &
 $\p_t+\mu\p_x+\big(\alpha_1v+q^2u\big)\p_v+$\newline{$\big(q^3u+\alpha_1w+p^3\big)\p_w, \ q^2=\alpha_2E(d_1,d),$}
 \newline{$q^3=\frac{a_2d_1-d}{d_1}\,q^2, \ p^3=\frac{d-a_2d_1}{d_1}\,q^2$}
 \\
\hline &&&\\
   13 &
$u(1-u)$  \newline{$uw$}
\newline{$-\frac{d}{d_1}uw$}
 &$d_2=d_3=d$ & $\p_t+\mu\p_x+\big(\alpha_1v+q^2u+h^2w+P(t,x)\big)\p_v+\big(q^3u+\alpha_2 w+p^3\big)\p_w, \
  h^2=\frac{d_1(\alpha_1-\alpha_2)}{d},$
 \newline{$q^2=\lf(\alpha_3e^{-\frac{d}{d_1}\,t}+\alpha_4\rg)E(d_1,d),$}
\newline{$q^3=-\frac{d}{d_1}\,q^2, \ p^3=\frac{\alpha_4d}{d_1}E(d_1,d)$}
 \\
\hline
\end{tabular}
\end{table}

\begin{remark}In Table~\ref{tab2} the function $P(t,x)$ is an arbitrary
solution of the linear diffusion equation $P_t =d_2P_{xx},$
$E(\delta_1,\delta_2)=\exp\lf[\mu\frac{\delta_2-
\delta_1}{2\delta_1\delta_2}\lf(x+\mu\frac{\delta_2-
\delta_1}{2\delta_1}\,t\rg)\rg],$ while $\alpha_i \ (i=1,2,3,4)$ and
$\mu$ are arbitrary constants.
\end{remark}
\begin{remark}  Some $Q$-conditional symmetry operators presented in the Table~\ref{tab2}
  are equivalent to the relevant Lie symmetry
  operators obtained in \cite{ch-dav-2017} provided   arbitrary
parameters satisfy additional restrictions. For example, operator
from Case~6 with $\alpha_1=-a_2d_3$ is the linear combination of Lie
symmetry operators $\p_t$ and $v\p_v+w\p_w$ (see Case~1 of Table~1
\cite{ch-dav-2017}).
 \end{remark}

\begin{remark} The HGF systems and relevant $Q$-conditional symmetry operators from Cases~4 and 5   of
Table~\ref{tab2}  can be reduced by the transformation \be\label{2-1} u^*=a_1w, \ v^*=-(u+a_1v), \ w^*=e^{-t}u, \ x^*=\frac{x}{\sqrt{d}} \ee to the subcases of Cases~7 and 5 from Table 2
\cite{ch-dav-2013}, respectively.
 \end{remark}

  \textbf{Sketch of the proof.} In order  to prove the theorem, one
needs to solve  the system of equations $3)$--$11)$ from (\ref{1-3})
under restrictions (\ref{1-8}) on the functions $C^k$  and taking
into account  that unknown functions have the structure (see  (\ref{1-5}))
 \be\nonumber\ba{l}\xi=\xi(t,x), \
\eta^1=r^1(t,x)u+q^1(t,x)v+h^1(t,x)w+p^1(t,x), \\
\eta^2=r^2(t,x)v+q^2(t,x)u+h^2(t,x)w+p^2(t,x), \\
\eta^3=r^3(t,x)w+q^3(t,x)u+h^3(t,x)v+p^3(t,x).\ea\ee

It turns out that essentially different  solutions of  equations $3)$--$11)$ from (\ref{1-3}) are obtained depending on
parameters $d_k$  and $a_i$ arising in the HGF system  (\ref{1-2}). All  such solutions are identified in what follows.

First of all, one notes that  equations  $3)$--$8)$ of system (\ref{1-3}) do not depend on
$u, \ v$ and $w$, and take the form \bea &&
 \label{3-8} (d_1-d_2)\,\xi q^1-2d_1d_2q^1_{x}=0, \
(d_1-d_3)\,\xi h^1-2d_1d_3h^1_{x}=0,
 \\
&& \label{3-9} (d_1-d_2)\,\xi q^2+2d_1d_2q^2_{x}=0,\ (d_2-d_3)\,\xi
h^2-2d_2d_3h^2_{x}=0, \\
&& \label{3-10} (d_1-d_3)\,\xi q^3+2d_1d_3q^3_{x}=0,
\  (d_2-d_3)\,\xi h^3+2d_2d_3h^3_{x}=0, \\
&& \label{3-11} d_1(\xi_{xx}-2r^1_{x})=\xi_t+2\xi\xi_x, \
d_2(\xi_{xx}-2r^2_{x})=\xi_t+2\xi\xi_x, \
d_3(\xi_{xx}-2r^3_{x})=\xi_t+2\xi\xi_x. \eea

On the other hand, equations $9)$--$11)$ of system (\ref{1-3}) can be splitted with
respect to the variables $u, v, w, uv,uw, vw, u^2, v^2$ and $w^2$.
As a result, one obtains the  system  \bea &&
\label{3-4} a_3h^1=0, \ d_3 h^1+\left(a_3 d_2+a_1 d_3\right) h^2=0,\  \left(a_5 d_2-a_1 a_2 d_3\right) h^3=0, \\
&& \label{3-5}  a_1\left(a_2 d_1-d_2\right) q^1=0, \ d_2 \left(a_4 d_1-2 d_3\right) h^1
-a_1 d_2 d_3 h^2-d_1 d_3 q^1=0,   \\
&& \label{3-6} d_2 \left(a_5 d_1-a_1 d_3\right) h^1-a_1 d_1 d_3 q^1=0,\\
&& \label{3-7} d_1 \left(a_4 d_2-a_2 d_3\right) h^3+d_2 \left(a_5 d_1-a_1 d_3\right) q^3=0,\
 \left(a_4 d_1-d_3\right) q^3=0, \\
&& \label{3-12} a_1 q^2+r^1+2 \xi_x=0, \ \left(a_2 d_1-d_2\right)
q^2-d_1 q^3=0, \\ && \label{3-12*} d_3 h^3+a_5 d_2 q^2+2 a_3 d_2
q^3+ a_4 d_2 \left(r^1+2 \xi_x\right)=0,\\ && \label{3-13} d_1
h^3-a_1 \left(2 a_2 d_1-d_2\right) q^2+a_1 d_1
q^3-a_2 d_1 \left(r^1+2 \xi_x\right)=0,   \\
&& \label{3-14} \left(a_2 d_1-2 d_2\right) q^1-a_1 d_2 \left(r^2+2 \xi_x\right)=0, \
a_1 h^3-a_2 q^1-a_1 a_2 \left(r^2+2 \xi_x\right)=0, \\
&& \label{3-15} \left(2 a_3 d_2+a_1 d_3\right) h^3+a_4 d_2 q^1+ a_5
d_2 \left(r^2+2\xi_x\right)=0,  \ a_4 h^1+a_5 h^2
=-a_3 \left(r^3+2 \xi_x\right), \\
&& \label{3-16} a_2 d_3 h^1-\left(a_5 d_2-2 a_1 a_2 d_3\right) h^2-
d_3 q^1=a_1 d_3 \left(r^3+2 \xi_x\right), \\
&& \label{3-17} \left(a_4 d_2-a_2 d_3\right) h^2+a_1 d_3 q^2+ d_3
\left(r^1-r^2\right)+d_3 \left(r^3+2 \xi_x\right)=0, \eea \bea &&
\label{3-18} d_1 r^1_{xx}-r^1_t- 2 r^1\xi_x+2\xi_x
+\left(\frac{d_1}{d_2}-1\right) q^1 q^2+
\left(\frac{d_1}{d_3}-1\right) h^1 q^3 -2
p^1-a_1 p^2=0,\\
&& \nonumber  d_2 r^2_{xx}-r^2_t- 2
r^2\xi_x+2a_2\xi_x+\left(\frac{d_2}{d_1}-1\right) q^1q^2+
\left(\frac{d_2}{d_3}-1\right) h^2 h^3-\\&&  \hskip2cm a_2 p^1-2 a_1 a_2 p^2+a_1 p^3=0,\\
&& \nonumber d_3 r^3_{xx}-r^3_t- 2
r^3\xi_x+2a_3\xi_x+\left(\frac{d_3}{d_1}-1\right) h^1 q^3+
\left(\frac{d_3}{d_2}-1\right) h^2 h^3-\\&&\label{3-20} \hskip2cm
a_4 p^1-a_5 p^2-2 a_3 p^3=0, \\  && d_1 q^1_{xx}-q^1_t-2 q^1
\xi_x+\left(1-\frac{a_2 d_1}{d_2}\right) q^1
+\left(\frac{d_1}{d_2}-1\right) q^1 r^2+\nonumber\\&&\label{3-21} \hskip2cm\left(\frac{d_1}{d_3} -1\right) h^1 h^3-a_1 p^1=0,\\
&& d_2 q^2_{xx}-q^2_t-2 q^2 \xi_x+\left(a_2-\frac{d_2}{d_1}\right)
q^2
+\left(\frac{d_2}{d_1}-1\right) q^2 r^1+\nonumber\\&&\label{3-22} \hskip2cm \left(\frac{d_2}{d_3}-1\right) h^2 q^3-a_2 p^2+p^3=0, \\
&& \label{3-23}d_3 q^3_{xx}-q^3_t-2 q^3 \xi_x +\left(a_3
-\frac{d_3}{d_1}\right) q^3 +\left(\frac{d_3}{d_1}-1\right) q^3
r^1+\left(\frac{d_3}{d_2}-1\right) h^3 q^2-a_4p^3=0,\eea \bea &&
\label{3-24} d_1 h^1_{xx}-h^1_t-2 h^1 \xi_x+\left(1-\frac{a_3
d_1}{d_3}\right) h^1+
\left(\frac{d_1}{d_3}-1\right) h^1 r^3+\left(\frac{d_1}{d_2}-1\right) h^2 q^1=0,\\
&& d_2 h^2_{xx}-h^2_t-2 h^2 \xi_x+\left(a_2 -\frac{a_3
d_2}{d_3}\right) h^2+
\left(\frac{d_2}{d_3}-1\right) h^2r^3+\nonumber\\&&\label{3-25} \hskip2cm\left(\frac{d_2}{d_1}-1\right) h^1 q^2+p^1+a_1 p^2=0, \\
&& d_3 h^3_{xx}-h^3_t-2 h^3 \xi_x+\left(a_3 -\frac{a_2
d_3}{d_2}\right) h^3+\nonumber\\&&\label{3-26} \hskip2cm
\left(\frac{d_3}{d_2}-1\right) h^3
r^2+\left(\frac{d_3}{d_1}-1\right) q^1 q^3-a_5p_3=0, \\ &&
\label{3-27}d_1 p^1_{xx}-p^1_t-2 p^1
\xi_x+p^1+\left(\frac{d_1}{d_2}-1\right) p^2 q^1+
\left(\frac{d_1}{d_3}-1\right) h^1 p^3=0, \\
&& \label{3-28}d_2 p^2_{xx}-p^2_t-2 p^2 \xi_x+a_2p^2+\left(\frac{d_2}{d_1}-1\right) p^1 q^2+
\left(\frac{d_2}{d_3}-1\right) h^2 p^3=0, \\
&& \label{3-29}d_3 p^3_{xx}-p^3_t-2 p^3
\xi_x+a_3p^3+\left(\frac{d_3}{d_1}-1\right) p^1
q^3+\left(\frac{d_3}{d_2}-1\right) h^3 p^2=0. \eea

Although the above system is very cumbersome, one is highly overdetermined
because the equation number is much larger than number of unknown functions $\xi, \ r^k, \ q^k, \ h^k$  and  $p^k$. Moreover, equations (\ref{3-4})--(\ref{3-7})  are algebraic (not PDEs\,!). It allows us to identify all inequivalent solutions of system (\ref{3-4})--(\ref{3-29}).

First of all, we note that the nonlinear system
(\ref{3-8})--(\ref{3-29}) in the case \[q^k=h^k=0 \ (k=1,2,3)\] is
reducible to the system of DEs for searching Lie symmetry operators.
All possible Lie symmetries were found in \cite{ch-dav-2017} (see Table 1 therein).

Now we observe that  a linear combination of equations (\ref{3-12})
and (\ref{3-13}) leads to   $d_3h^3=0,$ i.e. $h^3=0$. Moreover, {\it two
essentially different cases, $a_3\neq0$ and $a_3=0$,} follow from the
first equation of (\ref{3-4}).

Let us   examine in details  case
$a_3\neq0$. Having $h^3=0$ and assuming $a_3\neq0$,   equations
(\ref{3-4})--(\ref{3-7}), (\ref{3-12}) and (\ref{3-12*}) immediately
lead to \be\label{3-30}q^1=q^3=h^1=h^2=0.\ee Thus,  one needs to set $q^2\neq0$ in order to find
a non-Lie symmetry,
 therefore  equations (\ref{3-12}) and
(\ref{3-12*}) produce restrictions \be\label{3-31}
a_2=\frac{d_2}{d_1}, \ a_5=a_1a_4.\ee Moreover, using (\ref{3-30})
we obtain  $p^1=p^3=0$  from equations (\ref{3-21}), (\ref{3-23})
and (\ref{3-25}).

Hence, the system of DEs (\ref{3-8})--(\ref{3-29}) for finding
$Q$-conditional symmetries of the  HGF system (\ref{1-2}) with
$a_3\neq0$
 takes the form
 \bea && \label{3-33} r^2=r^3=-2 \xi_x, \\
&& \label{3-34} \xi_t-5d_2\,\xi_{xx}+2\xi\xi_x=0, \ \xi_t-5d_3\,\xi_{xx}+2\xi\xi_x=0,\\
&& \label{3-35} d_1(\xi_{xx}-2r^1_{x})=\xi_t+2\xi\xi_x,\\
 && \label{3-37}   a_1 q^2+r^1+2 \xi_x=0,\\
 && \label{3-38} d_1 r^1_{xx}-r^1_t- 2 r^1\xi_x+2\xi_x=0, \\
 && \label{3-39} d_2 r^2_{xx}-r^2_t- 2
r^2\xi_x+2\frac{d_2}{d_1}\,\xi_x=0, \\
 && \label{3-40} d_3 r^3_{xx}-r^3_t- 2
r^3\xi_x+2a_3\xi_x=0, \\
&& \label{3-41} (d_1-d_2)\, \xi q^2+2d_1d_2q^2_{x}=0, \\
&& \label{3-42} d_2 q^2_{xx}-q^2_t-2 q^2 \xi_x
+\left(\frac{d_2}{d_1}-1\right) q^2 r^1=0,  \\
&& \label{3-43} d_2 p^2_{xx}-p^2_t-2 p^2
\xi_x+\frac{d_2}{d_1}\,p^2=0, \  \  a_1 p^2=0.
  \eea

The corresponding $Q$-conditional symmetry has the form \be\nonumber
Q=\p_t+\xi\p_x+r^1u\p_u+(r^2v+q^2u+p^2)\p_v+r^3w\p_w.\ee

Let us integrate system (\ref{3-33})--(\ref{3-43}). Substituting
(\ref{3-33}) into (\ref{3-39}) and using the first equation of
(\ref{3-34}), one obtains the overdetermined system
\be\nonumber\xi_t+2\xi\xi_x-5d_2\,\xi_{xx}=0,\ \xi_{tx}+2\xi_x^2-d_2\,\xi_{xxx}+2\frac{d_2}{d_1}\,\xi_x=0.\ee
The general solution of this system is well-known
 (see (2.28) in \cite{a-h-b}):
\be\label{1*}\xi=\mu,\ee where $\mu$ is an arbitrary constant.
Having (\ref{1*}), we immediately obtain from equations
(\ref{3-33}), (\ref{3-35})--(\ref{3-38})
\[r^2=r^3=0, \ r^1_t=r^1_x=0, \ a_1 q^2+r^1=0.\]
If $a_1\neq0$ then $d_1=d_2$  (see (\ref{3-42})) and the general
solution of system (\ref{3-33})--(\ref{3-43}) leads to the operator
\[Q=\p_t+\mu\p_x+r^1\lf(u\p_u-\frac{1}{a_1}u\p_v\rg).\] This is nothing else but Lie symmetry operator (see Case~4 of Table~1
\cite{ch-dav-2017}).

If $a_1=0$ then the general solution of  system
(\ref{3-33})--(\ref{3-43}) has the form
\[\xi=\mu,\ r^1=r^2=r^3=0,\ q^1=q^3=h^k=0 \ (k=1,2,3), \ p^1=p^3=0,  \]
\[q^2=\lf(\alpha_1+\alpha_2e^{\frac{d_2}{d_1}\,t}\rg)\exp\lf(\mu^2\frac{(d_1-
d_2)^2}{4d_1^2d_2}\,t+\mu\frac{d_2-d_1}{2d_1d_2}\,x\rg),\]
\[p^2=-\alpha_2\exp\lf(\lf(\frac{d_2}{d_1}+\mu^2\frac{(d_1-
d_2)^2}{4d_1^2d_2}\rg)t+\mu\frac{d_2-d_1}{2d_1d_2}\,x\rg),\] where
$\alpha_1$ and $\alpha_2$ are arbitrary constants. Thus, Case~1 of
Table~\ref{tab2} is obtained and the case $a_3\neq0$ is completely
examined.

{\it The second generic case $a_3=0$} can be examined in a quite similar
way. The second equations of (\ref{3-4}) and (\ref{3-15})
\[ h^1+a_1h^2=0, \ a_4 h^1+a_5 h^2 =0\] lead to the restriction
$a_5=a_1a_4$ (otherwise $h^1=h^2=0 \Rightarrow q^1=q^2=q^3=0$, so
that only Lie symmetries can be derived).

Moreover, one notes from the first equation of (\ref{3-5}) that two
possibilities $a_1\neq0$ and $a_1=0$ should be analysed.

Let us assume that  $a_1\neq0$. Equations (\ref{3-5}), (\ref{3-7})
and (\ref{3-12}) lead to the restriction
$(a_2d_1-d_2)(a_4d_1-d_3)=0$ (otherwise we obtain
$h^1=h^2=q^1=q^2=q^3=0$). Thus, we need to consider two subcases:
\[\emph{(i)} \ a_2d_1-d_2\neq0 \Rightarrow a_4d_1=d_3; \quad
\emph{(ii)} \  a_2d_1=d_2.\]

Subcase $\emph{(i)}$. From the first equation of (\ref{3-5}) we find
$q^1=0$, and, as result, $p^1=p^3=0$ (see (\ref{3-21}) and
(\ref{3-26})). Integrating the first equation of (\ref{3-9}), we
find
\[q^2=\varphi(t)\exp\lf(\int
\frac{d_2-d_1}{2d_1d_2}\,\xi(t,x)dx\rg),\] were $\varphi(t)$ is
arbitrary smooth function, and $q^3=(a_2-\frac{d_2}{d_1})\,q^2$ from
the second equation of (\ref{3-12}). Substituting the function $q^3$
into the first equation of (\ref{3-10}), we have
\[(d_2-d_3)\,\varphi\,\xi=0.\]
If $\xi=0$, then $r^2=0$ (see the first equation (\ref{3-14})) and
\[q^2_x=q^3_x=h^1_x=h^2_x=0 \ (\texttt{see} \
(\ref{3-8})-(\ref{3-10})).
\]
Under the above equalities system (\ref{3-4})--(\ref{3-29})
essentially simplifies and takes the form
\bea && \label{3-51} q^2_t +
\left(1-\frac{d_1}{d_3}\right)\lf(a_2-\frac{d_2}{d_1}\rg)q^2h^2-p^2=0,\\
&&  \label{3-52}
\lf(a_2-\frac{d_2}{d_1}\rg)h^2_t+a_1\lf(a_2-\frac{d_2}{d_1}\rg)\left(\frac{d_3}{d_1}-1\right)
h^2 q^2+ a_1\frac{d_3}{d_1}\,p^2=0, \\  &&  \label{3-53}
q^2_t+\left(\frac{d_2}{d_1}-a_2\right) q^2
+a_1\left(\frac{d_2}{d_1}-1\right) \lf(q^2\rg)^2 +\lf(a_2-\frac{d_2}{d_1}\rg)\left(1-\frac{d_2}{d_3}\right) h^2 q^2=0, \\
&& \label{3-54}\lf(a_2-\frac{d_2}{d_1}\rg)q^2_t
+\frac{d_3}{d_1}\lf(a_2-\frac{d_2}{d_1}\rg) q^2
+a_1\lf(a_2-\frac{d_2}{d_1}\rg)\left(\frac{d_3}{d_1}-1\right)
\lf(q^2\rg)^2=0,\\
&& \label{3-55} h^2_t- h^2-
\lf(a_2-\frac{d_2}{d_1}\rg)\left(\frac{d_1}{d_3}-1\right)\lf(h^2\rg)^2=0\\
&& \label{3-56} h^2_t-a_2h^2-
\lf(a_2-\frac{d_2}{d_1}\rg)\left(\frac{d_2}{d_3}-1\right)\lf(h^2\rg)^2+a_1\left(\frac{d_2}{d_1}-1\right)
h^2 q^2- a_1 p^2=0, \eea and \be\nonumber\ba{l}q^1=r^2=p^1=p^3=0, \ p^2_t=0, \
a_2p^2=0, \\
r^1=-a_1q^2, \ q^3=\lf(a_2-\frac{d_2}{d_1}\rg)q^2,
\ h^1=-a_1h^2, \ r^3=\lf(a_2-\frac{d_2}{d_1}\rg)h^2. \ea\ee

Thus, we obtain the overdetermined nonlinear system of PDEs  with
two unknown functions $q^2$ and $h^2$ under the restriction
$\lf(q^2\rg)^2+\lf(h^2\rg)^2\neq0$ (otherwise  only Lie symmetries
can be derived). Note that this system is incompatible in the case
$q^2h^2\neq0.$ Integrating system (\ref{3-51})--(\ref{3-56}) for
$q^2=0, \ h^2\neq0$ and $h^2=0, \ q^2\neq0$, we obtain first and
second operators of Cases~2 of Table~\ref{tab2}, respectively.

If $\xi\neq0$, then only Lie symmetries can be obtained.

Subcase $\emph{(ii)}$ was examined in a similar way. As a result,
Cases~3, 4 and 5 of Table~\ref{tab2} have been derived.

 Thus, subcase $a_1\neq0$ is completely examined and
Cases 2--5 of Table~\ref{tab2} were obtained.

Finally, the possibility $a_1=0$ was examined. Because three
parameters $a_1, \ a_3$ and $a_5$ vanish, the system of DEs
(\ref{3-4})--(\ref{3-29}) simplifies essentially. As a result, Cases
6--13 of Table~\ref{tab2} were obtained in a straightforward way.

\textbf{The sketch  of the proof is now complete}.

\section{\bf Exact solutions of the HGF system} \label{sec:3}


If one compares the HGF systems  with the reaction terms arising
in Table~\ref{tab2} with its general form (\ref{1-2}) then it is clear
 that Cases 2--5 are the most interesting from the
applicability point of view. In fact, all the other cases of
Table~\ref{tab2} lead to the systems involving the autonomous Fisher
equation in the HGF system (\ref{1-2}). The autonomous Fisher
equation means that initial farmers $u$ does not interact with converted farmers $v$. So, it is  unlikely that such
systems can describe adequately the spread and interaction between
farmers and hunter-gatherers.

Here we study  in details the system with the reaction terms from Case~2 of
Table~\ref{tab2} because this system admits two $Q$-conditional
symmetries (i.e. possesses a wider symmetry) in contrast to those from Cases~3--5. Thus, we examine
the system
\begin{equation}\label{4-1}\begin{array}{l} \medskip  u_t = d_1 u_{xx}+u(1-u-a_1v), \\ \medskip v_t =
d_2 v_{xx}+ \frac{d_2-d_3}{d_1-d_3}\,v(1-u-a_1v)+uw+a_1vw, \\
w_t = d_3 w_{xx}-\frac{d_3}{d_1}uw-\frac{a_1d_3}{d_1}vw,
\end{array}\end{equation} where $a_1\neq0, \ d_1\neq d_2$ and $d_1\neq d_3.$

One can set $a_1=1$ without losing a generality because of  the
transformation $a_1v \rightarrow v, \ a_1w \rightarrow w$, hence
system (\ref{4-1}) and its operators take  the forms
\begin{equation}\label{4-2}\begin{array}{l} \medskip  u_t = d_1 u_{xx}+u(1-u-v),\\ \medskip v_t =
d_2 v_{xx}+ \frac{d_2-d_3}{d_1-d_3}\,v(1-u-v)+uw+vw, \\
w_t = d_3 w_{xx}-\frac{d_3}{d_1}uw-\frac{d_3}{d_1}vw,
\end{array}\end{equation}
\be\label{4-2*}\begin{array}{l} \medskip Q_1=\p_t-\frac{d_1}{d_1-d_2}\,w(\p_u-\p_v)-\frac{d_3}{d_1-d_3}\,w\p_w, \\
 Q_2=\p_t-\frac{d_3}{d_1-d_3}\,u\left(\p_u-\p_v\right)-\frac{(d_1-d_2)d_3^2}{d_1(d_1-d_3)^2}\,u\p_w.
 \end{array}\ee

It can be easily checked using Theorem~1~\cite{ch-dav-2017} that system
(\ref{4-2}) admits the trivial Lie symmetry allowing to search only
for plane wave solutions, in particular traveling fronts (see Section 4~\cite{ch-dav-2017}). Because
operators (\ref{4-2*}) present non-Lie symmetries, we construct here
exact solutions with more complicated structure.

It is well-known that using  any  $Q$-conditional symmetry, one can
reduce the given two-dimensional system of PDEs to a system of ODEs
via  the same procedure as for classical Lie symmetries. Thus, to
construct an ansatz corresponding to the operator $Q$, the system of
the linear first-order PDEs
 \be\label{4-3} Q\,(u)=0, \ Q\,(v)=0, \ Q\,(w)=0 \ee should be solved.

In the  case of the operator $Q_1$, system (\ref{4-3}) takes the form
\be\label{4-4}\begin{array}{l} u_t=-\frac{d_1}{d_1-d_2}\,w, \quad
v_t=\frac{d_1}{d_1-d_2}\,w, \quad
w_t=-\frac{d_3}{d_1-d_3}\,w.\end{array}\ee

Solving  system (\ref{4-4}), one obtains the ansatz
\begin{equation}\label{4-5}\begin{array}{l} \medskip u=\varphi_1(x)+ \frac{d_1(d_1-d_3)}{d_3(d_1-
d_2)}\,\varphi_3(x)\exp\lf(\frac{d_3}{d_3-d_1}\,t\rg),\\ \medskip
v=\varphi_2(x)-\frac{d_1(d_1-d_3)}{d_3(d_1-
d_2)}\,\varphi_3(x)\exp\lf(\frac{d_3}{d_3-d_1}\,t\rg),\\
w=\varphi_3(x)\exp\lf(\frac{d_3}{d_3-d_1}\,t\rg),
\end{array}\end{equation} where $\varphi_1, \ \varphi_2$ and
$\varphi_3$ are unknown functions.

In order to construct the reduced system, we substitute ansatz
(\ref{4-4}) into (\ref{4-2}). Making the relevant calculations one
arrives at the ODE system
\begin{equation}\label{4-6}\begin{array}{l} \medskip d_1\varphi_1''+\varphi_1\Big(1-\varphi_1-\varphi_2\Big)=0,\\ \medskip
d_2\varphi_2''+\frac{d_2-d_3}{d_1-d_3}\,\varphi_2\Big(1-\varphi_1-\varphi_2\Big)=0,\\
d_1\varphi_3''-\varphi_3\Big(\varphi_1+\varphi_2+\frac{d_1}{d_3-d_1}\Big)=0.
\end{array}\end{equation}

Using the same procedure for the operator $Q_2$, we obtain the ansatz
\begin{equation}\label{4-7}\begin{array}{l}  u= \varphi_1(x)\exp\lf(\frac{d_3}{d_3-d_1}\,t\rg),\\
v=\varphi_2(x)-\varphi_1(x)\exp\lf(\frac{d_3}{d_3-d_1}\,t\rg),\\
w=\varphi_3(x)+\frac{d_3(d_1-d_2)}{d_1(d_1-d_3)}\,\varphi_1(x)\exp\lf(\frac{d_3}{d_3-d_1}\,t\rg),
\end{array}\end{equation} and the ODE system
\begin{equation}\label{4-8}\begin{array}{l} d_1\varphi_1''-\varphi_1\lf(\frac{d_1}{d_3-d_1}+\varphi_2\rg)=0,\\
d_2\varphi_2''+\frac{d_2-d_3}{d_1-d_3}\,\varphi_2\lf(1-\varphi_2+\frac{d_1-d_3}{d_2-d_3}\,\varphi_3\rg)=0,\\
d_1\varphi_3''-\varphi_2\varphi_3=0.
\end{array}\end{equation}

The ODE systems (\ref{4-6}) and (\ref{4-8}) are nonlinear
systems, which are non-integrable. To the best of our knowledge, even particular  solutions of these ODE systems are unknown.
 Thus, we consider some special cases
allowing to construct their exact solutions.

Let us consider system (\ref{4-6}) with the additional restriction
$\varphi_2=1-\varphi_1,$ which essentially simplifies the system.
Thus,we immediately obtain
\[\varphi_1=\alpha x+\beta, \
\varphi_2=-\alpha x+1-\beta,\] where $\alpha$ and $\beta$ are
arbitrary constants. Because $\varphi_1+\varphi_1=1$, the third
equation of (\ref{4-6}) has the general solution
\[\varphi_3=c_1\sin\left(\sqrt{D}\,x\right)+c_2\cos\left(\sqrt{D}\,x\right), \ D=\frac{d_3}{d_1(d_1-d_3)}>0;\]
\[\varphi_3=c_1e^{\sqrt{|D|}\,x}+c_2e^{-\sqrt{|D|}\,x}, \ D<0\]
(hereafter $c_1$ and $c_2$ are arbitrary constants).
Substituting the functions $\varphi_1, \ \varphi_2$ and $\varphi_3$
into ansatz (\ref{4-5}),
we observe that \be\label{4-12}(u,v,w)
\rightarrow (\varphi_1,\varphi_2,0) \ \texttt{if}\ t \rightarrow
\infty, \ee provided $d_1>d_3$ (otherwise all the components tend to
infinity with time).

Because the asymptotic behavior  (\ref{4-12}) is plausible from the
applicability point of view, we concentrate ourselves on this case.
Assuming for a simplicity that $c_1=k(d_2-d_1)D>0, \ c_2=0$, $k>0$
and substituting the functions $\varphi_1, \ \varphi_2$ and
$\varphi_3$ into ansatz (\ref{4-5}), we obtain exact solution of the
HGF system (\ref{4-1})
\begin{equation}\label{4-11}\begin{array}{l}  u= -k\sin\left(\sqrt{D}\,x\right)\exp\lf(-d_1D\,t\rg)+\alpha x+\beta,\\
v=k\sin\left(\sqrt{D}\,x\right)\exp\lf(-d_1D\,t\rg)-\alpha x+1-\beta,\\
w=k(d_2-d_1)D\sin\left(\sqrt{D}\,x\right)\exp\lf(-d_1D\,t\rg).
\end{array}\end{equation}

\begin{figure}[t]
\begin{minipage}[t]{8cm}
  \quad  \quad \centerline{\includegraphics[width=8cm]{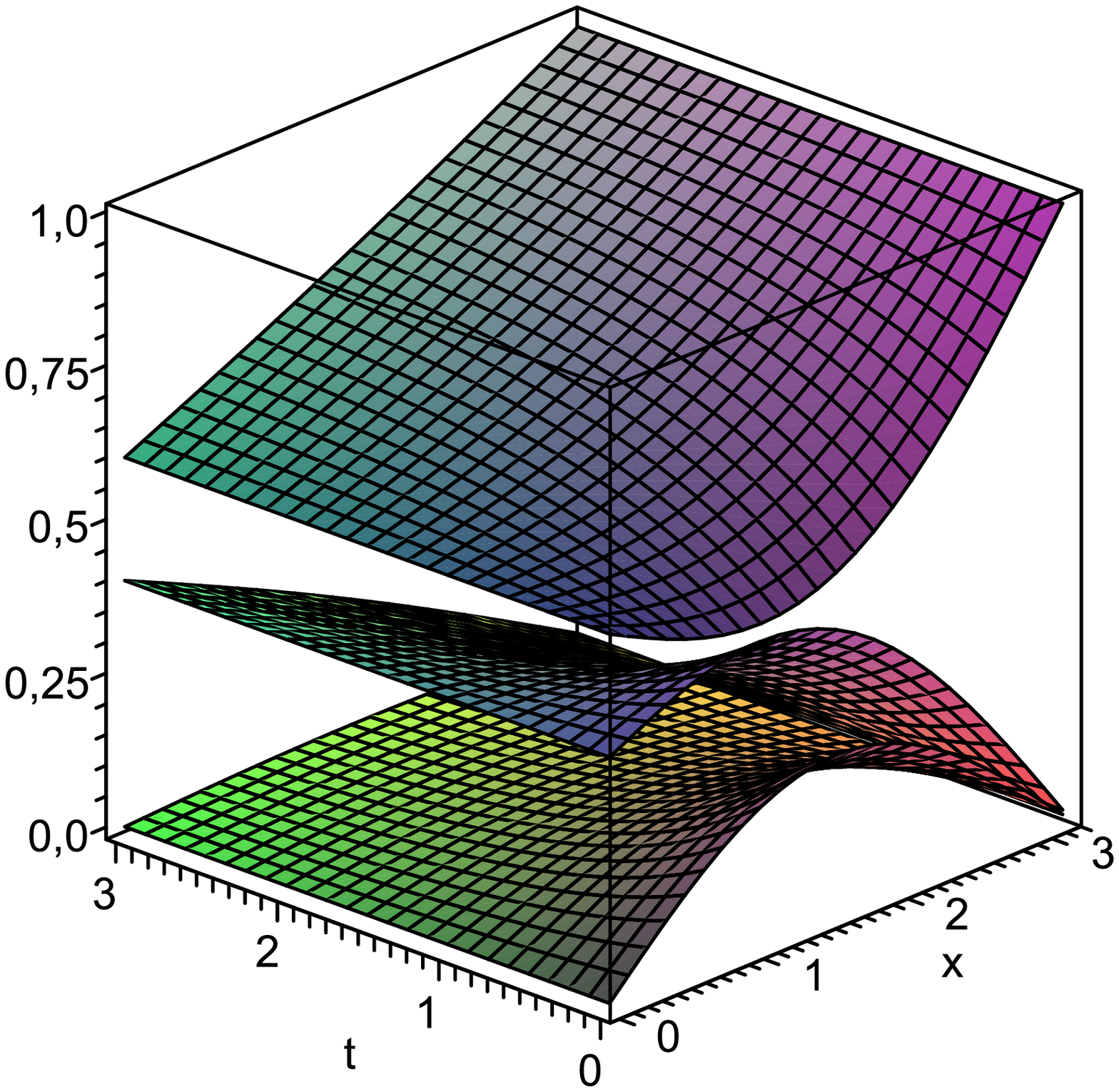}}
\end{minipage}
\begin{minipage}[t]{8cm}
  \quad  \quad \centerline{\includegraphics[width=8cm]{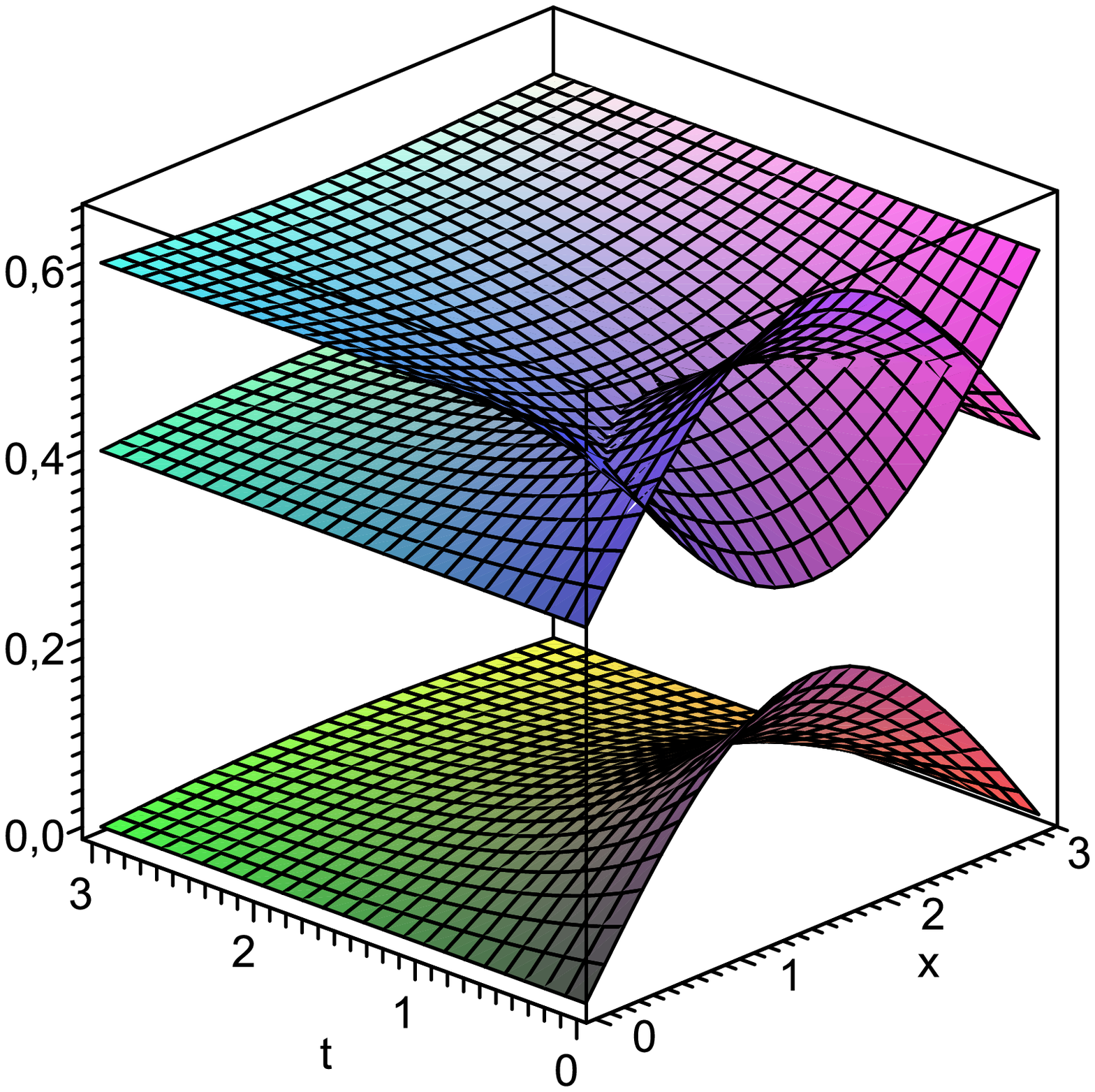}}
\end{minipage}
\center\caption{ Solution  (\ref{4-11})  of the HGF system
(\ref{4-2}) with $\alpha=1/8$ (left) and $\alpha=0$ (right). Other
parameters are: $k=1/4,  \ \beta=3/5, \ d_1=1, \ d_2=2, \ d_3=1/2.$
The upper,  middle  and lover surfaces represent  the functions $u$,  $v$,  and $w$, respectively. }
\label{f-1}
\end{figure}

 All the components of  solution  (\ref{4-11}) are
bounded and nonnegative in the domain \[\Omega=\lf\{ (t,x) \in (0,+
\infty )\times \lf(0,\frac{\pi}{{\sqrt{D}}}\rg)\rg\}\] if the
additional restrictions
\[\max\lf\{0,k-\frac{\alpha\pi}{2\sqrt{D}}\rg\}\leq\beta\leq1-\frac{\alpha\pi}{\sqrt{D}}, \ \texttt{if} \ \alpha\geq0,\]
\[\max\lf\{-\frac{\alpha\pi}{\sqrt{D}},k-\frac{\alpha\pi}{2\sqrt{D}}\rg\}\leq\beta\leq1, \ \texttt{if} \ \alpha<0,\]
take place. Notably, the  asymptotic behavior  (\ref{4-12}) takes the form
\be\label{4-12*}(u,v,w)
\rightarrow (\beta, 1-\beta, 0) \ \texttt{if}\ t \rightarrow
\infty \ee
in the border case $\alpha=0$. Here the point $(\beta, 1-\beta, 0)$ is nothing else but  a steady-state point of system (\ref{4-2}).

Solution (\ref{4-11}) has a clear biological interpretation and
describes such interaction between farmers and hunter-gatherers that
hunter-gatherers disappear  while the initial and converted farmers
coexist. Moreover, the population of initial farmers is increasing
with time, while the number of converted farmers is decreasing. An
examples of the solution are presented in Fig.~\ref{f-1}.

 We also point out that the components of solution (\ref{4-11})  obey the property $u+v=1$. Interestingly, the same property possess numerical solutions presented in Fig.2\cite{ao-sh-shige-96} (see curves for the components $F$ and $C$ excepting a vicinity of the point $ x=l$). We note that the curves representing $F$ and $C$ vanish at the point $ x=l$ because the zero Dirichlet conditions are used in \cite{ao-sh-shige-96}, hence the property $F+C=1$ is not valid in the vicinity of the point $ x=l$ (in contrast to our solution  (\ref{4-11})).

Now we turn to the reduced  system (\ref{4-8}) and  assume that the functions
$\varphi_2$ and $\varphi_3$ are linearly dependent. One notes that
 second and third equations of system (\ref{4-8}) coincide if
the restriction
\be\label{4-13*}\varphi_3=\frac{d_3(d_2-d_1)}{d_1(d_1-d_3)}\big(\varphi_2-1\big)\ee
takes place. Thus, we obtain the single  ODE \be \label{4-13}
d_1\varphi_2''=\varphi_2(\varphi_2-1).\ee

The general solution of the nonlinear  equation (\ref{4-13}) can be found in the
parametric  form \cite{pol-za}: \be\label{4-14}
\varphi_2=\pm\frac{3}{2}\,y, \ x=\sqrt{d_1}\int
\frac{dy}{\sqrt{c_1\pm y^3-y^2}}+c_2.\ee
Here the constant $c_2$  can be removed by
  applying the   space translation $x-c_2 \rightarrow x$.  Moreover,  we can obtain the
exact solution of equation (\ref{4-13}) in an explicit form for some correctly-specified values of  $c_1$. For instance,
solution (\ref{4-14}) with $c_1=0$ and $c_1=\frac{4}{27}$ has the
forms
\be\label{4-20}\varphi_2=\frac{3}{2}\lf(1+\tan^2\frac{x}{2\sqrt{d_1}}\rg)\ee
and
\be\label{4-21}\varphi_2=\frac{1}{2}\lf(-1+3\tanh^2\frac{x}{2\sqrt{d_1}}\rg),\ee
respectively.
Thus, the solution of the HGF system (\ref{4-2}) corresponding to
the function $\varphi_2$ from (\ref{4-20}) have the form
\begin{equation}\label{4-16}\begin{array}{l}  u= \varphi_1(x)\exp\lf(\frac{d_3}{d_3-d_1}\,t\rg),\\
v=\frac{3}{2}\lf(1+\tan^2\frac{x}{2\sqrt{d_1}}\rg)-\varphi_1(x)\exp\lf(\frac{d_3}{d_3-d_1}\,t\rg),\\
w=\frac{d_3(d_2-d_1)}{2d_1(d_1-d_3)}\lf(1+3\tan^2\frac{x}{2\sqrt{d_1}}\rg)+\frac{d_3(d_1-d_2)}{d_1(d_1-d_3)}\,\varphi_1(x)\exp\lf(\frac{d_3}{d_3-d_1}\,t\rg),
\end{array}\end{equation}
 where the function $\varphi_1(x)$ is an
arbitrary solution of the linear ODE
\be\label{4-15}d_1\varphi_1''-\varphi_1\lf(\frac{3d_3-d_1}{2(d_3-d_1)}+
\frac{3}{2}\,\tan^2\frac{x}{2\sqrt{d_1}}\rg)=0.\ee   It turns out  that
equation (\ref{4-15}) can be reduced to the hypergeometric equation:
\be\label{4-19}z(z-1)\psi''(z)-\lf(z+\frac{1}{2}\rg)\psi'(z)-
\frac{1}{2}\lf(1-\frac{3d_3-d_1}{d_3-d_1}\rg)\psi(z)=0\ee
by the substitution\cite{pol-za} \be\label{4-19*}
z=\sin^2\frac{x}{2\sqrt{d_1}}, \
\psi(z)=\cos^2\frac{x}{2\sqrt{d_1}}\,\varphi_1(x).\ee
Thus, the  general solution of the equation (\ref{4-19}) can be presented
in the form of the hypergeometric functions.

Equation (\ref{4-19}) is integrable in terms of elementary
functions provided  diffusivities  $d_1$ and $d_3$ have some  correctly-specified values.  For
example, this equation with $d_3=\frac{5}{9}d_1$ has the solution
$\psi(z)=(1-z)^{\frac{5}{2}}.$ Thus, using formulae (\ref{4-7}),
(\ref{4-13*}), (\ref{4-21}), (\ref{4-16}), (\ref{4-19*}) and transformation $x \
\rightarrow \sqrt{d_1}x$,  we obtain the exact solution
\begin{equation}\label{4-18}\begin{array}{l} \medskip u= c_1 \cos^3\frac{x}{2}\,\exp\lf(-\frac{5}{4}\,t\rg),\\ \medskip
v=\frac{3}{2}\lf(1+\tan^2\frac{x}{2}\rg)-c_1\cos^3\frac{x}{2}\,\exp\lf(-\frac{5}{4}\,t\rg),\\
w=\frac{5(d-1)}{8}\lf(1+3\tan^2\frac{x}{2}\rg)+c_1\frac{5(1-d)}{4}\,\cos^3\frac{x}{2}\,\exp\lf(-\frac{5}{4}\,t\rg)
\end{array}\end{equation} of the HGF system
\begin{equation}\label{4-17}\begin{array}{l} \medskip  u_t = u_{xx}+u(1-u-v),\\ \medskip v_t =
d v_{xx}+ \frac{9d-5}{4}\,v(1-u-v)+uw+vw, \\
w_t = \frac{5}{9} w_{xx}-\frac{5}{9}uw-\frac{5}{9}vw,
\end{array}\end{equation}
where $d=\frac{d_2}{d_1}$.

Now we observe that the exact solution (\ref{4-18}) possesses  the  asymptotic behavior
\be\nonumber(u,v,w)
\rightarrow \lf(0,\ \frac{3}{2}\lf(1+\tan^2\frac{x}{2}\rg), \ \frac{5(d-1)}{8}\lf(1+3\tan^2\frac{x}{2}\rg) \rg) \ \texttt{if}\ t \rightarrow
\infty. \ee
In contrast to the exact solution (\ref{4-11}), a possible interpretation of  (\ref{4-18}) says that
the initial  farmers disappear  while the  hunter-gatherers and converted farmers
coexist. Moreover, the limiting distribution of the the  hunter-gatherers and converted farmers is nonconstant.
The corresponding domain, in which all the components of  solution  (\ref{4-18}) are
bounded and nonnegative, can be easily specified if one sets $d>1$ and  $0<c_1 \leq\frac{1}{2}$. Of course, this scenario looks unrealistic, however we have shown mathematically that the HGF model with some  coefficients (for example as specified in (\ref{4-17})) admit a complete disappearance of  the initial  farmers.

Applying a similar  procedure for  the function $\varphi_2$  from  (\ref{4-21}),
 the exact solutions
\begin{equation}\label{4-22}\begin{array}{l} \medskip u= \cosh^3\frac{x}{2}\,e^{\frac{9}{4}\,t},\\ \medskip
v=\frac{1}{2}\lf(-1+3\tanh^2\frac{x}{2}\rg)-\cosh^3\frac{x}{2}\,e^{\frac{9}{4}\,t},\\
w=\frac{27(d-1)}{8}\lf(1-\tanh^2\frac{x}{2}\rg)+\frac{9(d-1)}{4}\,\cosh^3\frac{x}{2}\,e^{\frac{9}{4}\,t}
\end{array}\end{equation} and
\begin{equation}\label{4-24}\begin{array}{l} \medskip u= \sinh\frac{x}{2}\cosh^3\frac{x}{2}\,e^{4t},\\ \medskip
v=\frac{1}{2}\lf(-1+3\tanh^2\frac{x}{2}\rg)-\sinh\frac{x}{2}\cosh^3\frac{x}{2}\,e^{4t},\\
w=6(d-1)\lf(1-\tanh^2\frac{x}{2}\rg)+4(d-1)\sinh\frac{x}{2}\cosh^3\frac{x}{2}\,e^{4t}
\end{array}\end{equation}
of the HGF system (\ref{4-2}) have been constructed.
The exact solution (\ref{4-22}) is valid if $d_1=1, \ d_2=d$ and $d_3=\frac{9}{5}$, while   (\ref{4-24}) is valid if $d_1=1, \ d_2=d$
and $d_3=\frac{4}{3}$.
We note that solutions (\ref{4-22}) and (\ref{4-24}) are growing unboundedly with time, therefore  their biological interpretation is  questionable.



\section{\bf Conclusions} \label{sec:4}

In this paper, $Q$-conditional (nonclassical) symmetry
  of the non-linear three-component  system (\ref{1-2}) used for describing the spread of  farmers into a
 region occupied by hunter-gatherers was under study.
 The main result is formulated in Theorem 1 and it says that there are exactly 13 inequivalent  systems of the form (\ref{1-2}) admitting  $Q$-conditional  symmetry operators of the form (\ref{1-6}). The result is rather unusual because the  corresponding nonlinear system of DEs (see (\ref{1-3})-(\ref{1-8})) was fully integrated without any additional restrictions. For example, the nonlinear system of DEs corresponding to the three-component DLV system was not solved in \cite{ch-dav-2013} but only under the restriction that  the  symmetries in question are {\it $Q$-conditional symmetries of the first type} (see the definition in \cite{ch-2010}). A natural question arises: Why conditional symmetry of a more complicated system can be easier identified than the DLV system (\ref{1-1*})? We have the following  hypothesis:  systems  involving  PDEs with the same structure (the DLV system is a typical example) possesses a wider conditional symmetry comparing with those involving equations with different structures. Roughly speaking, a symmetric structure of PDE systems leads to a wider symmetry.

 It is interesting that two systems among  the HGF systems admitting $Q$-conditional  symmetry are reducible to the three-component DLV systems. It occurs in Cases 4 and 5 of Table~\ref{tab2}, that the local substitution (\ref{2-1})
  reduces the corresponding systems to those arising in Cases~7 and 5 of Table 2 \cite{ch-dav-2013}.
 However, all other systems of the form (\ref{1-2}) with the reaction terms listed in Table~\ref{tab2}  are not reducible to any DLV system. It means that  the relevant $Q$-conditional symmetries are indeed new and cannot derived from those presented in \cite{ch-dav-2013}.

Each $Q$-conditional symmetry listed in Table~\ref{tab2} can be applied for reduction of the relevant HGF system to a system of ODEs  and search for  exact solutions. Here the symmetries listed in Case 2  of Table~\ref{tab2} were examined in order to find exact solutions of system (\ref{4-2}) because the latter is the most interesting among others from both mathematical and applicability point of view. As a result several solutions in explicit form   were derived (see formulae (\ref{4-11}), (\ref{4-18}), (\ref{4-22}) and (\ref{4-24})). The most interesting among them is the exact solution (\ref{4-11}), which  describes  plausible scenarios of interaction  between the three populations and possesses (with correctly-specified parameters) similar properties to  numerical solutions presented in the pioneering work \cite{ao-sh-shige-96}. In particular, the solution predicts the scenario when
hunter-gatherers disappear  while the initial and converted farmers
coexist and their densities tend with time (see formula (\ref{4-12*})) to the steady-state point of system (\ref{4-2}).

Finally, we point out that the following problem is still open: to find  Q-conditional symmetries of the HGF system in the so-called  no-go case, i.e. to  construct operators of the form~(\ref{1-6**}). Our experience in the case of two-component reaction-diffusion systems \cite{ch-dav-2015} says   that some progress can be done in this direction if one applies the definition of  Q-conditional symmetry of the first type \cite{ch-2010}. Another possibility is to use the method of  heir equations introduced in \cite{nucci-96}.

\end{document}